\documentclass[aps,physrev,twocolumn,superscriptaddress,nofootinbib]{revtex4-2}
\usepackage{subfigure}
\usepackage[version=4]{mhchem}
\usepackage{wasysym}
\usepackage[separate-uncertainty,retain-explicit-plus,per-mode=symbol,binary-units]{siunitx}
\usepackage{array,mathtools,amssymb,dcolumn}
\usepackage{tikz}
\usepackage{multirow}
\usepackage{afterpage}
\usepackage{lineno}
\usepackage{paralist}
\usepackage{listings}
\usepackage{array}
\usepackage{cancel}
\usepackage{xspace}
\UseRawInputEncoding
\usepackage{stmaryrd}
\usepackage[version=4]{mhchem}

\newcommand{\FPrompt}{\mbox{$F_{\text{prompt}}$}}

\begin{document}

%\preprint{}

\title{Search for gravitational wave coincident neutrinos in DEAP-3600}

% Define institutions
\newcommand{\Alberta}{Department of Physics, University of Alberta, Edmonton, Alberta, T6G 2R3, Canada}
\newcommand{\AstroCeNT}{AstroCeNT, Nicolaus Copernicus Astronomical Center, Polish Academy of Sciences, Rektorska 4, 00-614 Warsaw, Poland}
\newcommand{\BHSU}{School of Natural Sciences, Black Hills State University, Spearfish, SD 57799, USA}
\newcommand{\Cagliari}{Physics Department, Universit\`a degli Studi di Cagliari, Cagliari 09042, Italy}
\newcommand{\UCR}{Center for Experimental Cosmology \& Instrumentation, Department of Physics and Astronomy, University of California, Riverside, CA 92521, USA}
\newcommand{\CNL}{Canadian Nuclear Laboratories, Chalk River, Ontario, K0J 1J0, Canada}
\newcommand{\Carleton}{Department of Physics, Carleton University, Ottawa, Ontario, K1S 5B6, Canada}
\newcommand{\CIEMAT}{Centro de Investigaciones Energ\'eticas, Medioambientales y Tecnol\'ogicas, Madrid 28040, Spain}
\newcommand{\Houston}{Department of Physics, University of Houston, Houston, TX 77204, USA}
\newcommand{\INFNCagliari}{INFN Cagliari, Cagliari 09042, Italy}
\newcommand{\INFNNapoli}{INFN Napoli, Napoli 80126, Italy}
\newcommand{\Mainz}{Institut f\"ur Kernphysik, Johannes Gutenberg-Universit\"at Mainz, 55128 Mainz, Germany}
\newcommand{\LNGS}{INFN Laboratori Nazionali del Gran Sasso, Assergi (AQ) 67100, Italy}
\newcommand{\LU}{School of Natural Sciences, Laurentian University, Sudbury, Ontario, P3E 2C6, Canada}
\newcommand{\LBNL}{Nuclear Science Division, Lawrence Berkeley National Laboratory, Berkeley, CA 94720, USA}
\newcommand{\UNAM}{Instituto de F\'isica, Universidad Nacional Aut\'onoma de M\'exico, A.\,P.~20-364, Ciudad de M\'exico~01000, Mexico}
\newcommand{\Napoli}{Physics Department, Universit\`a degli Studi ``Federico II'' di Napoli, Napoli 80126, Italy}
\newcommand{\Capodimonte}{Astronomical Observatory of Capodimonte, Salita Moiariello 16, I-80131 Napoli, Italy}
\newcommand{\MEPhI}{National Research Nuclear University MEPhI, Moscow 115409, Russia}
\newcommand{\Oxford}{Department of Physics, University of Oxford, Oxford, OX1 3PU, United Kingdom}
\newcommand{\IIPA}{International Institute for Particle Astrophysics, Polish Academy of Sciences,  Bartycka 18, 00-716 Warsaw, Poland}
\newcommand{\Princeton}{Physics Department, Princeton University, Princeton, NJ 08544, USA}
\newcommand{\Queens}{Department of Physics, Engineering Physics, and Astronomy, Queen's University, Kingston, Ontario, K7L 3N6, Canada}
\newcommand{\RHUL}{Royal Holloway University London, Egham Hill, Egham, Surrey TW20 0EX, United Kingdom}
\newcommand{\RAL}{Rutherford Appleton Laboratory, Harwell Oxford, Didcot OX11 0QX, United Kingdom}
\newcommand{\SL}{SNOLAB, Lively, Ontario, P3Y 1M3, Canada}
\newcommand{\Sussex}{University of Sussex, Sussex House, Brighton, East Sussex BN1 9RH, United Kingdom}
\newcommand{\TRIUMF}{TRIUMF, Vancouver, British Columbia, V6T 2A3, Canada}
\newcommand{\TUD}{Institut f\"ur Kern und Teilchenphysik, Technische Universit\"at Dresden, 01069 Dresden, Germany}
\newcommand{\TUM}{Department of Physics, Technische Universit\"at M\"unchen, 80333 Munich, Germany}
\newcommand{\MI}{Arthur B. McDonald Canadian Astroparticle Physics Research Institute, Queen's University, Kingston, ON, K7L 3N6, Canada}

% DEAP alphabetic order of institutions
\affiliation{\Alberta}
\affiliation{\AstroCeNT}
\affiliation{\BHSU}
\affiliation{\Cagliari}
\affiliation{\UCR}
\affiliation{\CNL}
\affiliation{\Carleton}
\affiliation{\CIEMAT}
\affiliation{\Houston}
\affiliation{\INFNCagliari}
\affiliation{\INFNNapoli}
\affiliation{\Mainz}
\affiliation{\LU}
\affiliation{\UNAM}
\affiliation{\Capodimonte}
\affiliation{\Napoli}
\affiliation{\MEPhI}
\affiliation{\Oxford}
\affiliation{\IIPA}
\affiliation{\Queens}
\affiliation{\RHUL}
\affiliation{\RAL}
\affiliation{\SL}
\affiliation{\TRIUMF}
\affiliation{\TUD}
\affiliation{\TUM}
\affiliation{\MI}

% Set authors
\author{P.~Adhikari}\affiliation{\Carleton}
\author{P.-A.~Amaudruz}\affiliation{\TRIUMF}
\author{D.\,J.~Auty}\affiliation{\Alberta}
\author{M.~Batygov}\affiliation{\LU}
\author{B.~Beltran}\affiliation{\Alberta}
\author{S.~Belayneh~Habtemariam}\affiliation{\AstroCeNT}
\author{M.\,A.~Bigentini}\affiliation{\Carleton}\affiliation{\MI}
\author{C.\,E.~Bina}\affiliation{\Alberta}\affiliation{\MI}
\author{W.~Bonivento}\affiliation{\INFNCagliari}
\author{M.\,G.~Boulay}\affiliation{\Carleton}
% ...

\collaboration{DEAP Collaboration}
\altaffiliation{deap-papers@snolab.ca}

\date{\today}

\begin{abstract}
We searched DEAP-3600 data for neutrino signals coincident with gravitational waves from compact object mergers observed by the LIGO, Virgo, and KAGRA collaborations (LVK). While mergers involving neutron stars are expected to produce neutrinos, these have never been detected. The DEAP-3600 experiment is sensitive to electron neutrinos as recently observed through charged current neutrino absorption on its $^{40}$Ar target. Between 2016 and 2020, the detector recorded data during seven compact merger events involving neutron stars that were observed gravitationally by LVK. Upon examination, there is no statistically significant evidence of neutrinos from any of the mergers analyzed. 
Exclusion limits are set on the electron neutrino fluence, e.g. from GW170817 in the mono-energetic model at 6.8$\times10^{12}~\text{cm$^{-2}$}$ (1.8$\times10^{10}~\text{cm$^{-2}$}$) for 10~MeV (100~MeV) neutrinos.
\end{abstract}

%\keywords{}

\maketitle

\renewcommand{\thefootnote}{\roman{footnote}}

\section{Introduction}
Since the first detection of gravitational waves (GWs) in 2015 by the LIGO and Virgo collaborations~\cite{firstGW}, a precise record of many compact binary coalescences (CBCs) has been available to physicists. The third GW data catalog (GWTC-3.0~\cite{LIGO-3}) included two binary neutron star (NSNS) mergers, five neutron star-black hole (NSBH) mergers, and 83 binary black hole (BHBH) mergers. This presents an opportunity to search for other signals from these mergers, as the GW information can be used to establish a limited time window in which to expect them. For example, in 2017 the Swope Supernova Survey used a GW signal to help identify light from a kilonova that followed after a NSNS merger~\cite{swope}.

A vast body of nuclear astrophysics literature concurs that neutrinos will be produced in and ejected from CBCs involving neutron stars (NSs) or black holes (BHs) \cite{burns,kimuraNSNS,foucart,ardevol}. As far back as 1989, and encouraged by the detection of neutrinos from a supernova in 1987 \cite{sn1987nu}, proposals have been made for what kind of equipment would be needed to accomplish this \cite{eichler}. As such, many detector experiments have carried out searches for neutrinos from CBCs (including ANTARES~\cite{antares}, Borexino~\cite{borexino}, IceCube~\cite{icecube}, KM3NeT~\cite{km3net}, Super-Kamiokande~\cite{superk}, and XENON1T \cite{xenon}) with no successful detections so far. This study continues that work in DEAP-3600, looking only at mergers involving neutron stars.

The DEAP-3600 experiment has collected data since 2016 with the goal of detecting dark matter and other exotic particles in the universe using a single-phase liquid argon (LAr) scintillation detector \cite{deap3600_design}. The detector was principally designed to search for the dark matter candidates Weakly Interacting Massive Particles (WIMPs) with minimal backgrounds. Owing to the low backgrounds and dense target, it was recently found to also be able to observe interactions with solar neutrinos through charged current neutrino absorption on $^{40}$Ar nuclei~\cite{nuAbsPaper}. According to LVK data~\cite{LIGO-3,LIGO-1,LIGO-2,LIGO-2.1}, nine mergers likely involving a NS occurred while DEAP-3600 was taking data over the course of its ``second fill'' from November 2016 to March 2020. In this work, the data taken during seven of these mergers are analyzed in comparison with empirical background models to search for GW-coincident neutrino interactions.

\section{Signal}
The high density environment within a CBC allows for multiple theoretical mechanisms of neutrino production, including nuclear decays and electron-positron interactions~\cite{foucart,ardevol,eichler}. For decades, astrophysical calculations have attempted to predict the characteristics of these neutrinos~\cite{eichler,berezinsky}. Due to their low interaction cross section, neutrinos can reach Earth after traveling for many megaparsecs from a merger. With enough exposure then, a detector could record a neutrino interaction correlated in time with the merger.

The emission of neutrinos from a merger is expected to occur broadly in two populations, high energy (on the GeV-TeV scale) and low energy (on the keV-MeV scale)~\cite{burns}. The former may contribute to the polar jet flux, and may be detectable alongside electromagnetic signals from short $\gamma$-ray bursts~\cite{kimuraNSNS}. The latter, referred to as thermal neutrinos, are expected in abundance from the outer layers of the merger~\cite{tamborra}. Since it has not yet yet been observed, the energy spectrum, flux, and flavor of neutrino emission from mergers involving a NS is not known. Models common in nature and/or in other astrophysical processes are often used as placeholders to allow for reasonable limits to be explored in the absence of a detected signal. 

Four such models are used to test theoretical limits on electron neutrino fluence in this analysis: flat, power law, supernova-like, and mono-energetic. These have each been used in prior searches: a flat spectrum by Super-Kamiokande~\cite{superk}, a power law spectrum by ANTARES, IceCube, and Super-Kamiokoande~\cite{antares,icecube,superk-170817}, a supernova-like spectrum by Borexino~\cite{borexino}, and a mono-energetic model by Borexino, Super-Kamiokande, and XENON1T~\cite{xenon,superk-170817,borexino}. The flat spectrum predicts a uniform distribution of neutrinos with respect to energy. Though this model is not common in nature it is nonetheless useful for exploring the case in which the neutrino emission is sparse enough to hide energy dependence. Power law spectra, by contrast, are commonly found in nature~\cite{pwrLaws}. This analysis tested an inverse energy squared model. A supernova-like spectrum, normalized from Equation~2.4 of Huang et. al.~\cite{superSpectrum}, was tested because it is a neutrino signal that has been detected on Earth before~\cite{sn1987nu}. Though the spectrum of neutrinos from NS mergers has a very different origin, this is a reasonable starting point as it is well understood and unique in shape to the previous two models. Finally, the mono-energetic model allows for examining detector sensitivity at any energy.

The neutrino flux from a CBC is expected to peak most significantly during the 1000~second window uniformly surrounding the arrival of its GW on Earth~\cite{baret}. While this window is primarily meant to encapsulate neutrinos of an energy much higher than those best studied by DEAP-3600, it has been adopted by other experiments for use with keV-MeV energy scale neutrinos~\cite{borexino,superk,xenon}. 
This narrow coincidence window is used to select data with the highest chance of containing merger neutrino events, and it also allows the use of sideband data as an empirical background model.

\section{Detector \& Event Reconstruction}
DEAP-3600 holds (3269 $\pm~24$) kg~\cite{Adhikari:2023ces} of liquid argon in a spherical acrylic vessel surrounded by 255 photomultiplier tubes (PMTs). Acrylic light guides connect the exterior of the acrylic vessel to the front of each PMT. The vessel and PMTs are sealed inside a stainless steel shell which is suspended from above by a cylindrical ``neck'' and hangs inside a roughly eight-meter diameter water tank. An additional 48 PMTs line the outside of the steel shell, enabling the water tank to serve as a muon veto. The detector is housed underground in Cube Hall at SNOLAB in Ontario, Canada.

Particles traversing the detector produce ultraviolet scintillation light in the LAr target. This ultraviolet light is wavelength shifted to 420~nm visible light by a 3~$\mu$m-thick layer of tetraphenyl butadiene (TPB) on the inner surface of the acrylic vessel. Visible photons are detected by the inward-looking PMTs as charge signals that are then converted into a total number of photoelectrons (PE) collected in each event. The detector response is linear for events depositing less than 2.5~MeV of visible energy, with a light yield of 7.1~PE/keV before afterpulsing removal. This light yield is calibrated run-by-run ($\sim$daily)  using the known 1.46~MeV $\gamma$-ray from $^{40}$K.

\begin{figure}[t]
    \centering
    \includegraphics[width=0.95\linewidth]{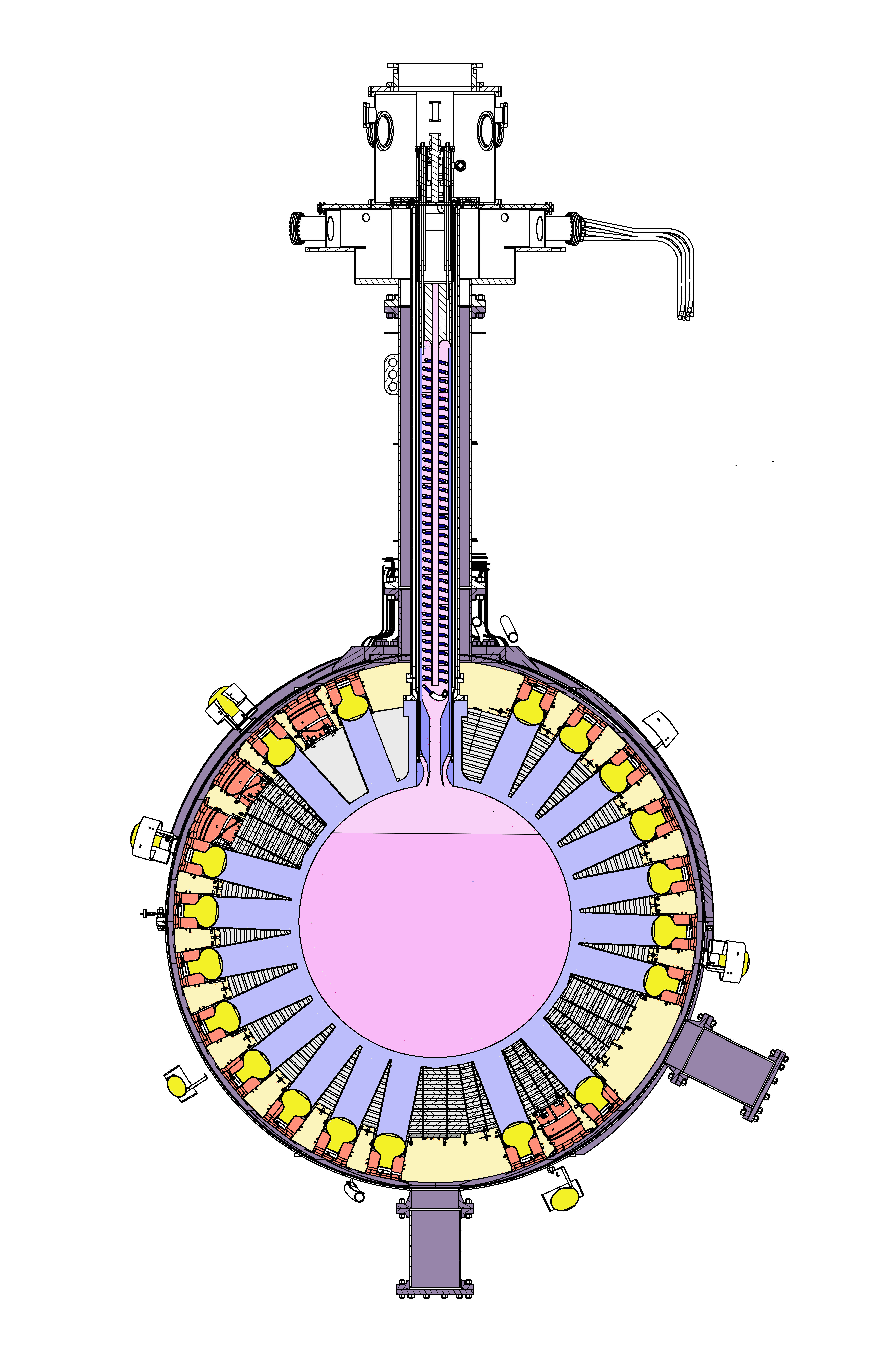}
    \caption{A cross sectional diagram of the DEAP-3600 detector with the LAr volume in dark pink, acrylic vessel and light guides in blue, PMTs in yellow, and stainless steel sphere and neck in dark purple.}
    \label{fig:detector}
\end{figure}

Above 2.5~MeV of visible energy this relationship becomes less accurate due to saturation effects. To account for this, a quadratic model is used with the form
\begin{equation}
    \text{PE}(E_{\text{vis}})= \text{A}\cdot E_{\text{vis}}^{2} + \text{B}\cdot E_{\text{vis}} + \text{C} 
    \label{eq:CarlsModel}
\end{equation}
where $E_{\text{vis}}$ is the visible energy measured in MeV, 
\mbox{$\text{A} = -19.51$~PE/MeV$^2$}, 
\mbox{$\text{B} = 6854$~PE/MeV}, and 
\mbox{$\text{C} = 549$~PE}~\cite{Rethmeier:2021oez}. 
This model was derived using fits to americium-beryllium (AmBe) calibration source data in which (n, $\gamma$) peaks are observable up to 10~MeV. An evaluation of the uncertainty propagation involved with extending the model to higher energies was performed~\cite{AndrewThesis}; this allowed the DEAP-3600 solar neutrino search to extend up to 13~MeV, i.e. about 86350~PE~\cite{nuAbsPaper}, without introducing significant uncertainty. 
The energy resolution $\sigma$ in PE of the model is given by
\begin{equation}
    \sigma = \sqrt{(1+\sigma^2_{\text{PE}})\cdot \text{PE} + \sigma^2_{\text{rel, LY}}\cdot \text{PE}^{2}}
\end{equation}
where $\sigma^2_{\text{PE}}= 2.97$~PE is a resolution scaling factor that accounts for effects
such as the Fano factor and PE counting noise,
and $\sigma^2_{\text{rel, LY}}= 3.0\times10^{-5}$ accounts for the relative light yield variance.

For this analysis, charged current electron neutrino absorption is used as the interaction channel. This is the neutrino channel most visible above the detector backgrounds, and it was observed in LAr for the first time in DEAP~\cite{nuAbsPaper}. At lower energies, electron neutrino absorption~\cite{PhysRevD.34.2088,PhysRevC.58.3677,PhysRevC.80.055501} results in bound states and follows the process
\begin{equation}
    \begin{array}{l}
        \nu_e~ +~ \ce{^40Ar}~ \rightarrow~ \ce{^40K^*}~ +~ e^-\\
        \\
         \ce{^40K^*}~ \rightarrow~ \ce{^40K}~ +~ \gamma_n \\ 
    \end{array}  
\end{equation}
where $\ce{^40K^*}$ is an excited state of $\ce{^40K}$ and $\gamma_n$ is a $\gamma$-ray cascade. The effective threshold for electron neutrino absorption is 3.83~MeV, representing the transition from $^{40}$Ar to the 2.33~MeV state of $^{40}$K. Up to roughly 10~MeV, the interaction in argon results in de-excitation photons from the decay of \ce{^40K^*} and the emission of a prompt electron. The $\gamma$-ray emission can come in multiple cascades, and approximately one third of transitions to a high enough energy state of $^{40}$K will decay through a 1.64~MeV metastable state with a lifetime of 480~ns~\cite{GARDINER2021108123,Gardiner_PhysRevC.103.044604}. 
Decay through this metastable state thus results in a delayed coincidence signature. At higher energies, unbound final states become possible which can include a neutron, $\alpha$-particle, or both. The total energy $E_{\nu}$ deposited by a neutrino in the detector can be represented by
\begin{equation}
    E_{\nu} = E_e + E_{\gamma} + Q = E_{\rm vis} + 1.5~\text{MeV}
\end{equation}
for a bound final state where $E_e$ is the prompt electron energy, $E_{\gamma}$ is the total energy from de-excitation $\gamma$-rays, and $Q = 1.5~\text{MeV}$ is the Q-value.

\section{Event Selection \& Dataset}
\label{Section:CutsData}
% Table is defined in this section to allow proper float location in the PDF.
\begin{table*}[t]
    \centering
    \caption{List of all mergers involving a NS detected by LVK during the second fill of DEAP-3600 which corresponded with usable data. Merger distance is given in LVK data releases 1-3 (including 2.1)~\cite{LIGO-1,LIGO-2,LIGO-2.1,LIGO-3}. 
    Dead time fraction is the percentage of the 1000~second coincidence window's data lost to overlapping events or routine triggers. 
    The longer ID of GW190426\_152155 differentiates it from a second merger recorded on the same day.}
    \begin{tabular}{l c c c c c}
        \hline
        \textbf{Merger ID} & \textbf{Merger Type} & \textbf{Date (UTC)} & \textbf{Distance [Mpc]} & \textbf{Dead Time Fraction} & \textbf{Background Model} \\
        \hline
        GW170817 & NSNS & 08/17/2017 & 40 & 3.77\% & AmBe-in-MiniCLEAN \\
        %\hline
        GW190425 & NSNS & 04/25/2019 & 160 & 4.70\% & $^{22}$Na-in-DEAP \\
        %\hline
        GW190426\_152155 & NSBH & 04/26/2019 & 370 & 3.59\% & Source-Free \\
        %\hline
        GW190917 & NSBH & 09/17/2019 & 720 & 3.62\% & Source-Free \\
        %\hline
        GW200105 & NSBH & 01/05/2020 & 270 & 3.53\% & Source-Free \\
        %\hline
        GW200115 & NSBH & 01/15/2020 & 290 & 3.63\% & Source-Free \\
        %\hline
        GW200210 & NSBH & 02/10/2020 & 940 & 3.63\% & Source-Free \\
        \hline
    \end{tabular}
    \label{fullRunlist}
\end{table*}
Candidate signal events are distinguished from backgrounds in DEAP-3600 data using a series of event selection cuts, similar to the selection used in~\cite{nuAbsPaper}. Electronic recoil events originating in the LAr volume are selected where the fraction of total light collected in the neck veto PMTs is less than 0.5\%, a maximum of 20\% of the total light is detected in the PMT that observed the maximum amount, the time of the main peak in the waveform is well-aligned with the hardware trigger, and there are fewer than four early pulses before the main peak.
Up to two peaks are allowed in the waveform, to account for the metastable decay state, while rejecting higher-order pile-up.
Finally, events are rejected if they occur within an asymmetric \mbox{[-0.15, +1.35]}~ms window of a muon veto trigger; this is an anti-coincidence cut.

The parameter \FPrompt~describes the fraction of PE from an event emitted in a fixed prompt window. It is defined as
\begin{equation}
    F_{\rm prompt} = \frac{\sum^{150{\rm ns}}_{t=-28{\rm ns}}PE(t)}{\sum^{10\mu{\rm s}}_{t=-28{\rm ns}}PE(t)}
\end{equation}
where $PE(t)$ is the number of photoelectrons collected in the given summation window. \FPrompt~is used to distinguish nuclear recoils from electronic recoils, a method known as pulse shape discrimination (PSD) in LAr~\cite{adhikari_pulse-shape_2021}. Electron neutrino absorption events result in an electronic recoil with an \FPrompt~value between 0.1 and 0.35.

The region of interest (ROI) for this analysis therefore includes events in the aforementioned 1000~second search window around a GW signal, that survive all cuts and satisfy $0.1 <$~\FPrompt~$< 0.35$ and PE~$>16,885$. This PE value corresponds roughly to 2.4~MeV. An upper limit on PE is not placed in this analysis as there is no theoretical limit to the emission energy of neutrinos from mergers involving a NS. 

\begin{figure}[t]
    \centering
    \includegraphics[width=1.0\linewidth]{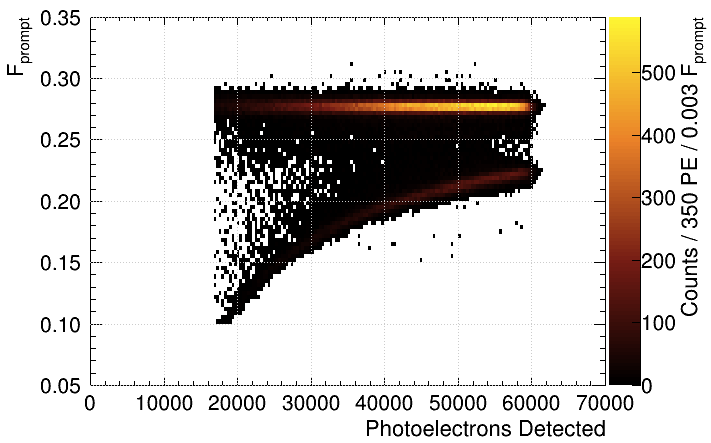}
    \caption{Histogram of \FPrompt~vs. PE for 1.5$\times10^5$ simulated neutrinos up to 10~MeV with all event selection cuts applied. One third of the events are drawn randomly from each of the three spectral signal models (excluding the mono-energetic model). The signal acceptance as a function of energy is found by discretizing the results from the flat spectrum simulation. The two distinct populations represent the more common (upper stripe) and metastable (lower stripe) decay paths.}
    \label{fig:simScatter}
\end{figure}

\begin{figure}
    \centering
    \includegraphics[width=1.0\linewidth]{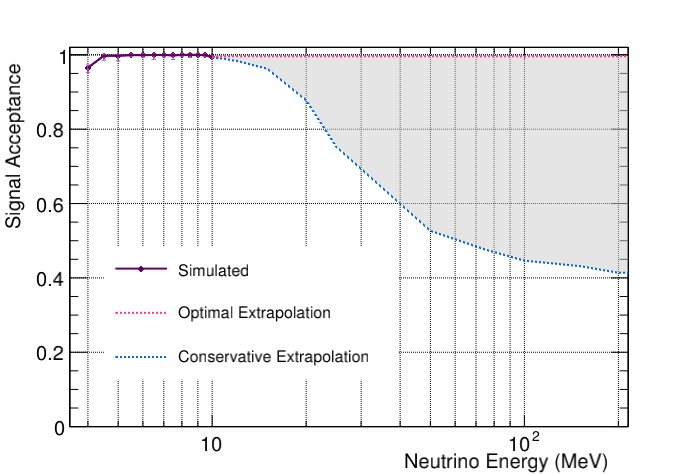}
    \caption{Signal acceptance as a function of neutrino energy for the event selection used in this analysis.  
    Below 10~MeV the acceptance is calculated from simulations. 
    Shown above 10~MeV are the optimal extrapolation and the conservative extrapolation described in the text.}
    \label{fig:sigAcc}
\end{figure}

Simulations of electron neutrino absorption in the detector using \texttt{RAT}~\cite{RAT} and \texttt{MARLEY}~\cite{GARDINER2021108123} show that for the lower portion of the ROI (up to 10~MeV), the event selection yields an average signal acceptance of $98 \pm 1$\%.
Simulated signal events are shown in Figure~\ref{fig:simScatter}. 
Most of the small acceptance loss comes from rejecting events that are reconstructed below the lower PE limit. Starting at 10~MeV, unbound final states become increasingly more probable, reaching a nearly 60\% combined branching ratio by 100~MeV. These states include an additional neutron and/or alpha particle and could not be simulated with full optical information in the DEAP infrastructure due to high memory demand. 
Thus, in this analysis two approaches are taken, the first being the optimal case where unbound final states are not rejected by the event selection cuts: in this approach the signal acceptance above 10~MeV is taken to be the average signal acceptance below 10~MeV.
At the other extreme, we consider a conservative extrapolation assuming that all unbound final state events are rejected by the cuts, therefore multiplying the optimal signal acceptance by the probability of bound final states.
The resulting signal acceptance as a function of energy is shown in Figure~\ref{fig:sigAcc}.

Nine mergers involving a NS occurred during the second data taking period from November 2016 to March 2020. Table~\ref{fullRunlist} shows a list of the seven of these mergers that occurred during standard data collection and resulted in usable data. Data are considered usable if they pass quality checks, were taken with all detector systems running, and contain the entire uninterrupted 1000-second signal search window from a merger. 
The percentage of each 1000-second window that is lost due to dead time is also given in Table~\ref{fullRunlist}. GW190814 is rejected because DEAP data are available for less than half of the 1000~second window, and GW191219\_163120 because the muon veto was offline.

This leaves seven mergers that occurred during runs passing all quality cuts: two NSNS and five NSBH. Merger GW190425 has atypical properties, but is nonetheless believed to be NSNS~\cite{gw190425}. Merger GW200210 could also possibly be BHBH~\cite{gw200210}, but as this is uncertain it is considered NSBH for this analysis. Two of the NSBH mergers (GW190426\_152155 and GW200105) are ``marginal'' and not ``confident'' observations by LVK, but are included here regardless. The two NSNS mergers occurred during atypical DEAP runs, with an additional background source present. 
Merger GW170817 occurred while a neighboring detector, MiniCLEAN~\cite{miniclean}, had an AmBe source deployed for calibration: this is a neutron source with an energy ranging up to 11~MeV, so there is a potential extra background source from (n,$\gamma$) reactions present in the data taken by DEAP during this merger. Merger GW190425 occurred while DEAP had its $^{22}$Na calibration source deployed. The peak of the $\gamma$-ray emission of $^{22}$Na is below the lower PE limit of this analysis, however the upper tail of the spectrum is not. Furthermore, during this type of calibration run only 10\% of the data recorded is saved (prescaling) so as to not overwhelm the data acquisition system. Thus, during merger GW190425 there was both a potential extra background source present, and less data available during the corresponding 1000-second signal window.

\section{Backgrounds}
\label{Section:BG}
Empirical background models are created using sideband data taken outside all coincidence windows. The same quality criteria applied to the data taken during mergers is applied to these sidebands. In addition, only data taken while LVK was operating are considered. While LVK is offline there is currently no other way to know if a merger has occurred. Applying this restriction therefore minimizes the probability of using data taken during an unobserved merger and adding signal events to a background model.

As expected, higher background is observed in the sideband data corresponding to the two runs when sources were present.
Data taken while the MiniCLEAN AmBe source was deployed shows an event rate between 5 to 10 MeV roughly 28\% higher than in any of the five source-free samples:
consistent with possible AmBe neutron background interactions.
In data taken while the DEAP $^{22}$Na source was deployed, the factor of 10 prescale in data results in the comparison being too inconsistent to identify potential unique backgrounds. 
Therefore, the two runs with sources present are analyzed separately from regular ``source-free'' data,
requiring the preparation of three different background models. 
The amount of each relevant type of data that passed quality checks is given in Table~\ref{tab:modelData}. These three portions of data are used to make each background model; the ``AmBe-in-MiniCLEAN'' model is used in the analyis of data taken during GW170817, the ``$^{22}$Na-in-DEAP'' model in the case of GW190425, and the source-free model corresponds to the five NSBH mergers.

\begin{table}[b]
    \centering
    \caption{The total amount of each type of data that passed quality checks and was therefore used in a background model.}
    \begin{tabular}{c c}
        \hline
        \textbf{Data Type} & \textbf{Live time [hours]} \\
        \hline
        MiniCLEAN AmBe Deployed & 47.06 \\
        DEAP $^{22}$Na Deployed & 45.67 \\
        Normal Physics Runs & 5867.75 \\
        \hline
    \end{tabular}
    \label{tab:modelData}
\end{table}

\begin{figure}[h]
    \centering
    \includegraphics[width=1.0\linewidth]{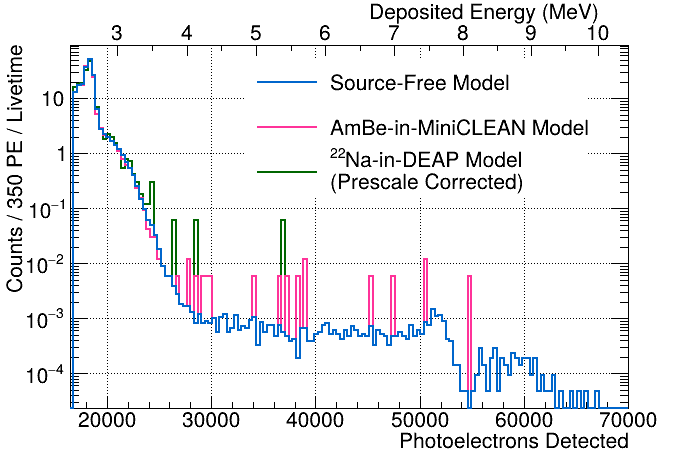}
    \caption{The three background models scaled to their respective merger livetimes (or 1000~seconds for the source-free model). 
    The effects of prescaling on the $^{22}$Na-in-DEAP model are corrected here to enable the comparison.
    Gaps in the two source-present models have been filled with a minimum corresponding to the source-free model expectation. There are no events in any background model above 70,000 PE.}
    \label{fig:BGsource}
\end{figure}

After applying the event selection to the data used for each model, the remaining backgrounds are consistent with the DEAP electromagnetic background model described in Ref.~\cite{PhysRevD.100.072009}. Figure~\ref{fig:BGsource} shows all three background models scaled to the livetime of the related merger data (or 1000~seconds for the source-free model). 
To remove under-fluctuations otherwise present in the AmBe-in-MiniCLEAN and $^{22}$Na-in-DEAP models, in each bin of the scaled PE histogram, the source-free model expectation is taken as the minimum number of events expected.
Then, the expected number of events in the coincidence window for a given merger is found by scaling the appropriate background model to the livetime of the coincidence window.

In order of prevalence, the dominant sources in the background models are $^{208}$Tl $\gamma$-rays, $^{42}$Ar-$^{42}$K $\beta$-decays, and (n,$\gamma$) reactions. Detector components including the stainless steel shell and the PMT glass contain trace amounts of $^{232}$Th, which decays into $^{208}$Tl, whose $\gamma$-ray emission peaks near 2.6~MeV of visible energy. The LAr within DEAP is distilled from the atmosphere, and as such it contains traces of cosmogenically produced $^{42}$Ar. This decays to $^{42}$K which in turn undergoes a $\beta$-decay with an emission spectrum that extends just above that of the $^{208}$Tl $\gamma$-rays~\cite{PhysRevD.100.072009}. Finally, many detector materials are neutron or $\alpha$-particle sources. Whether from spontaneous fission or ($\alpha$,n) reactions, emitted neutrons can capture on other detector materials to create an (n,$\gamma$) emission spectrum that reaches energies up to 10~MeV.

The statistical uncertainty on the background expectation $n_i$ for each bin of a given model is taken to be the Poissonian $\sqrt{n_i}$. These uncertainty values, as with the expectations, are scaled appropriately to the livetime of their search windows. The results are analyzed with the ROI divided into three regions for which a statistical uncertainty is also calculated. These regions, R1 for high backgrounds, R2 for low backgrounds, and R3 for zero backgrounds, help to better conceptualize where events are observed. Details of the region boundaries are shown in Table~\ref{tab:regions}. 
R2 is selected such that the total number of events expected within it during 1000~seconds is below 0.5. R1 contains the remaining range down to the bottom of the ROI. 
In R3, where zero events are seen in any of the runs from which the background models are derived, a 90\% confidence interval upper limit scaled by the ratio of the merger data livetime to the model livetime is used.

\begin{table}[h]
    \centering
    \caption{The boundaries in PE and incoming neutrino energy for each of the three background regions defined.}
    \begin{tabular}{c c c}
        \hline
         \textbf{Background} & \textbf{PE Range} & \textbf{Approx. E$_{\nu}$} \\
         \textbf{Region} && \textbf{Range [MeV]} \\
         \hline
         R1 (high background) & 16,885 -- 23,100 & 4.0 -- 4.8 \\
         R2 (low background)\ \ & 23,100 -- 67,200 & 4.8 -- 11.5 \\
         R3 (zero background) & $>$ 67,200 & $>$ 11.5 \\
         \hline
    \end{tabular}
    \label{tab:regions}
\end{table}

The main contributor to systematic uncertainty in this analysis is the effect of uncertainties on the light yield in each run used to build the models. 
Variations in the light yield measured in DEAP-3600~\cite{deap_39Ar_halflife} are corrected in the data by taking calibrated light yield measurements from every run and adjusting all PE values accordingly. 
To evaluate the systematic uncertainty, the PE correction factor for each background model for every run is varied by one standard deviation. 
The uncertainty on signal acceptance shown below 10~MeV in Figure~\ref{fig:sigAcc} comes from the same source: events crossing the PE threshold depending on the light yield value.
The systematic uncertainty is calculated on both the total number of events expected and on the number of events expected in each region (R1-R3) for each merger.

\section{Results}
\label{Section:Results}
\begin{figure*}[h!tb]
 \centering
\includegraphics[width=0.49\linewidth]{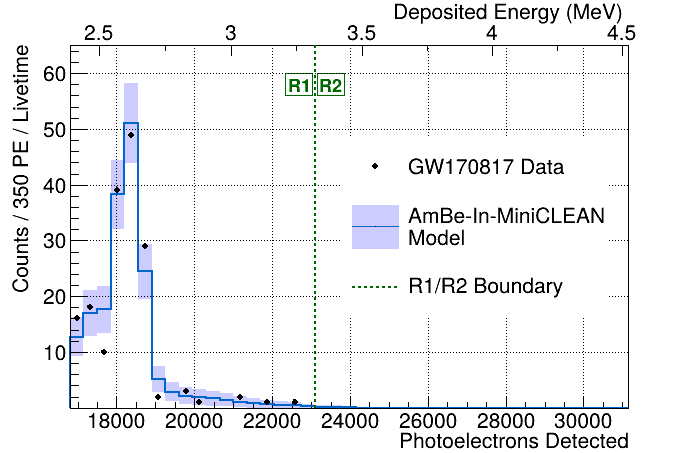} \includegraphics[width=0.49\linewidth]{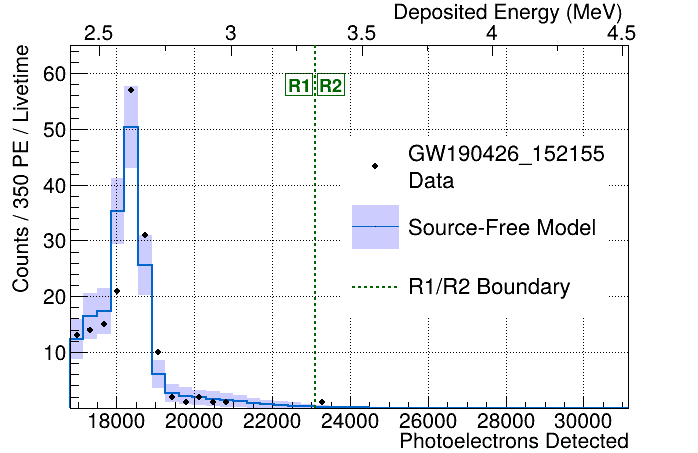} \includegraphics[width=0.49\linewidth]{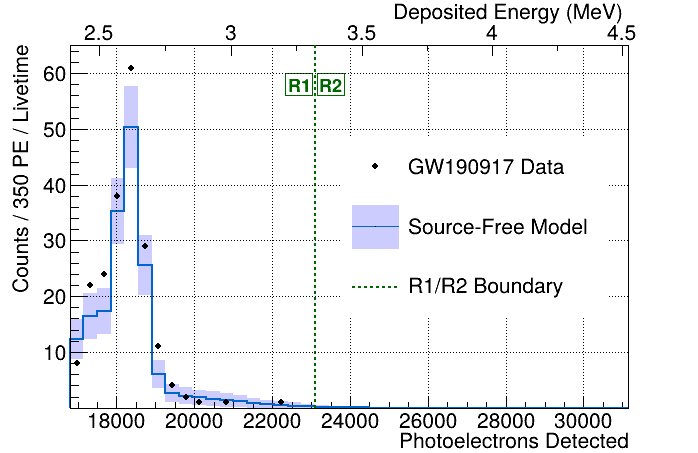} \includegraphics[width=0.49\linewidth]{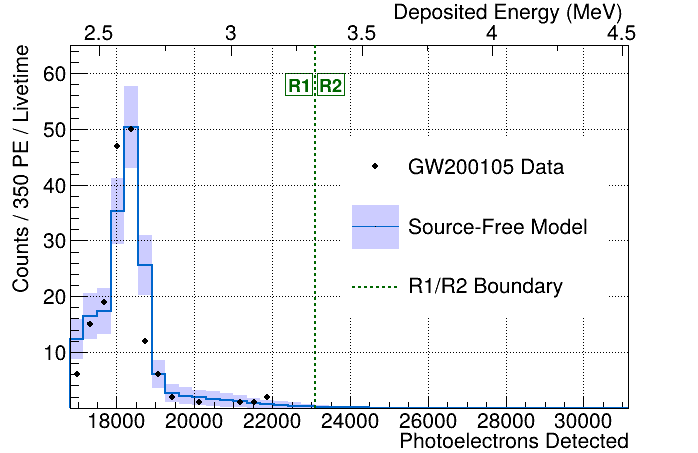} \includegraphics[width=0.49\linewidth]{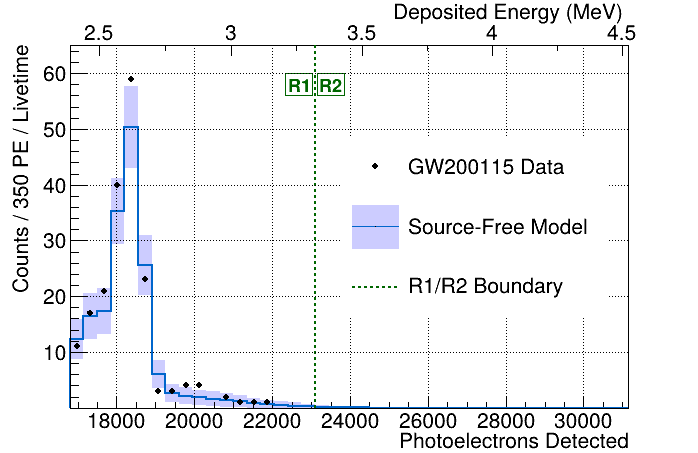} \includegraphics[width=0.49\linewidth]{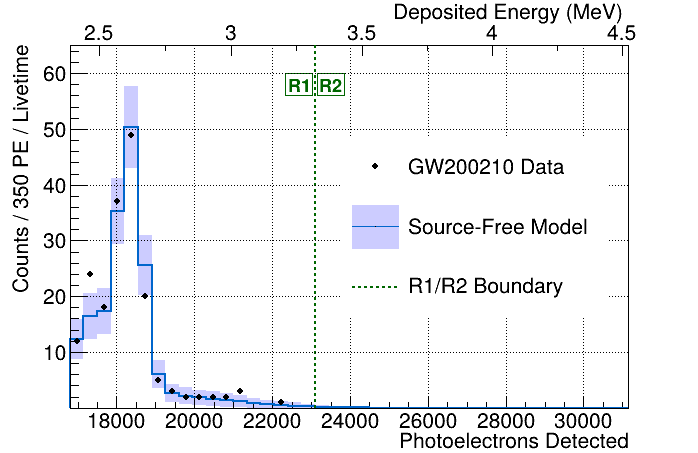}  \caption{Histograms of the events in the ROI during six of the GW signal windows, with the expected background counts from the source-free model (AmBe-in-MiniCLEAN model for GW170817) shown in blue.
 }
 \label{fig:dataVsBG}
\end{figure*}

\begin{figure}[htb]
    \centering
    \includegraphics[width=1.0\linewidth]{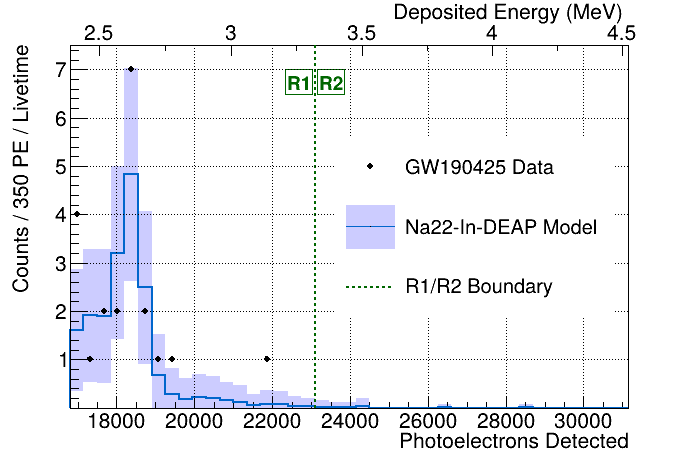}
    \caption{Histogram of the events in the ROI during the GW190425 signal window, with the expected background counts from the $^{22}$Na-in-DEAP model shown in blue, taking into account prescaled data acquisition in this run.}
    \label{fig:dataVsBG_GW190425}
\end{figure}

Table~\ref{tab:BGexpect} shows the expected number of backgrounds during each merger's livetime according to the appropriate background model, the corresponding uncertainty, and the observed number of events. The number of events in R1 for each merger is consistent with the background expectations to within 2$\sigma$. Only one event across all seven mergers appears outside of R1: this observed event, in R2 for GW190426\_152155, yields a p-value of 0.376 given the background expectation of (0.47 $^{+0.69}_{-0.47}$)~events. Figure~\ref{fig:dataVsBG} shows the data from six of the mergers against the background model prediction, and  
Figure~\ref{fig:dataVsBG_GW190425} shows the same for merger GW190425 which took place during a prescaled run.

\renewcommand{\arraystretch}{1.5} % Default value: 1
\begin{table*}[t]
    \centering
    \caption{Expected and observed number of events in the data during each merger, along with the $p$-value for the observation against the background-only model. The uncertainties given are the quadrature sum of the statistical and systematic uncertainties. For R3, only the scaled upper limit of the 90\% confidence interval is given as the expectation.
    }
    \begin{tabular}{l|c c c|c c c|c c c}
        \hline
        \textbf{Merger ID} & \textbf{R1 Exp.} & \textbf{R1 Obs.} & \textbf{$p$-value} & \textbf{R2 Exp.} & \textbf{R2 Obs.} & \textbf{$p$-value} & \textbf{R3 Exp.} & \textbf{R3 Obs.} & \textbf{$p$-value} \\
        \hline
        GW170817 & 180 $\pm$14 & 171 & 0.262 & 0.47 $^{+0.69}_{-0.47}$ & 0 & 0.625 & $<$ 0.014 & 0 & 0.986 \\
        GW190425 & 18.1 $\pm$4.3 & 21 & 0.206 & 0.09 $^{+0.30}_{-0.09}$ & 0 & 0.913 & $<$ 0.014 & 0 & 0.986 \\
        GW190426\_152155 & 177 $\pm$13 & 168 & 0.264 & 0.47 $^{+0.69}_{-0.47}$ & 1 & 0.376 & $<1.1\times10^{-4}$ & 0 & 0.999 \\
        GW190917 & 177 $\pm$13 & 202 & 0.035 & 0.47 $^{+0.69}_{-0.47}$ & 0 & 0.624 & $<1.1\times10^{-4}$ & 0 & 0.999 \\
        GW200105 & 177 $\pm$13 & 162 & 0.135 & 0.47 $^{+0.69}_{-0.47}$ & 0 & 0.624 & $<1.1\times10^{-4}$ & 0 & 0.999 \\
        GW200115 & 177 $\pm$13 & 190 & 0.172 & 0.47 $^{+0.69}_{-0.47}$ & 0 & 0.624 & $<1.1\times10^{-4}$ & 0 & 0.999 \\
        GW200210 & 177 $\pm$13 & 180 & 0.419 & 0.47 $^{+0.69}_{-0.47}$ & 0 & 0.624 & $<1.1\times10^{-4}$ & 0 & 0.999 \\
        \hline
    \end{tabular}
    \label{tab:BGexpect}
\end{table*}
\renewcommand{\arraystretch}{1} % Default value: 1

\begin{figure}[!h]
    \centering
    \includegraphics[width=0.95\linewidth]{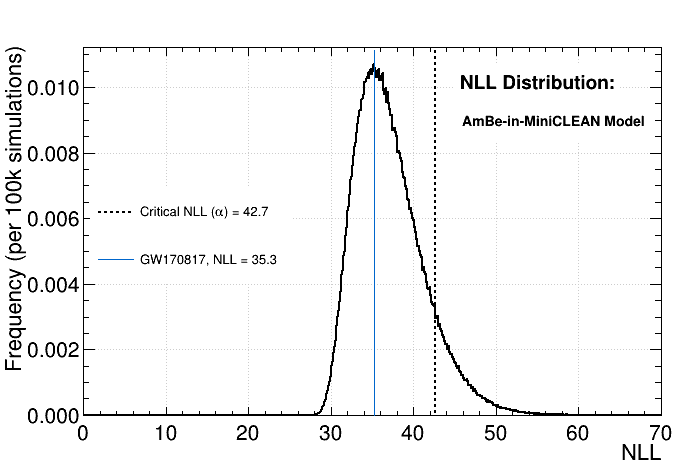}
    \label{fig:NLLD_mini}
    \includegraphics[width=0.95\linewidth]{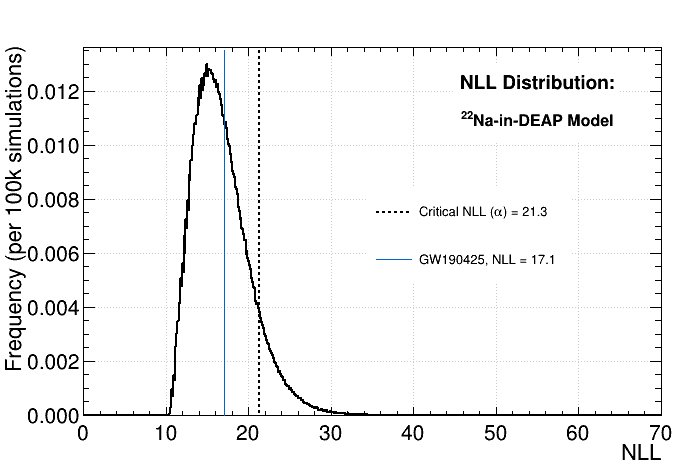}
    \label{fig:NLLD_na22}
    \includegraphics[width=0.95\linewidth]{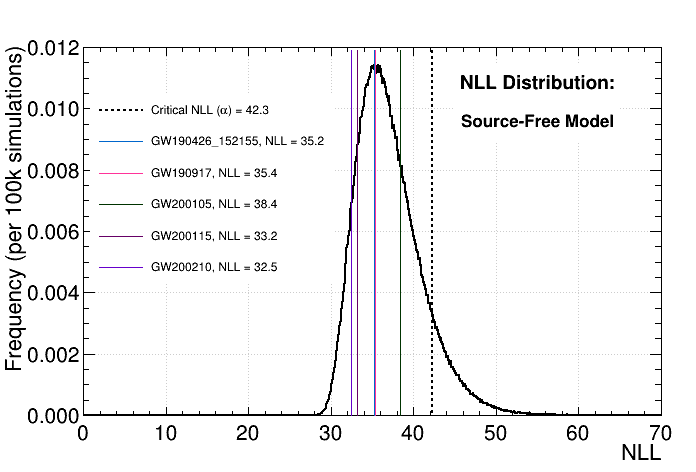}
    \caption{The NLL distributions for the AmBe-in-MiniCLEAN (top), $^{22}$Na-in-DEAP (middle), and source-free (bottom) models with their 90\% confidence level critical NLL values ($\alpha$) shown. Also shown are the NLL values of each merger's data against the relevant background model.
    }
    \label{fig:NLLD_uni}
\end{figure}

A negative log likelihood (NLL) goodness-of-fit test is used to more robustly assess the consistency of the data taken during each merger with the corresponding background model. The NLL of two distributions, an expectation and an observation, is computed as:
\begin{equation}
    \text{NLL} = -\ln\mathcal{L} = - \sum^n_{i=1} \ln\left(\frac{(\lambda_i)^{k_i}e^{-\lambda_i}}{k_i!}\right)
\end{equation}
where $n$ is the number of bins in both distributions, $\lambda_i$ is the expected number of events in bin~$i$, and $k_i$ is the observed number. 
This NLL is minimized when the two distributions are highly similar. 
The critical region boundary $\alpha$ used for this test is the NLL value for which there is less than a 10\% chance to see that value or higher, based on 
NLL distributions from pseudo-data drawn from the background-only hypothesis.
For each background model, the background-only NLL distribution is obtained by running $10^5$ pseudo-experiments, varying the model according to its uncertainties each time.
The resulting distributions are shown in Figure~\ref{fig:NLLD_uni}.

To test the background-only hypothesis, the NLL values for the data observed during each merger versus its background model are calculated and compared to the appropriate $\alpha$. These are shown in Table~\ref{tab:NLL} and visualized against their distributions in Figure~\ref{fig:NLLD_uni}. 
All seven observed NLL values fall below their respective $\alpha$, meaning that the data collected during all seven mergers are consistent with the background-only hypothesis.

Exclusion limits on the fluence $\Phi$ (time-integrated flux) of electron neutrinos that reached the detector from each merger are drawn using the formula
\begin{equation}
    \Phi_{\rm limit}(E_i) = \frac{N_{\rm limit}(E_i)}{n_{t}\varepsilon(E_i)\phi(E_i)\sigma(E_i)}
\end{equation}
where $E_i$ is the neutrino energy at the center of bin $i$, $N_{\rm limit}$ is the 90\% confidence interval upper limit on the number of signal events in the bin corresponding to $E_i$, $n_t$~is the number of target argon atoms in the detector, $\varepsilon$~is the signal acceptance at $E_i$, $\phi$~is the signal flux probability from the normalized model spectrum at $E_i$, and $\sigma$~is the interaction cross section at $E_i$ calculated using the implementation from Bhattacharya et al.~\cite{PhysRevC.80.055501} as detailed in Ref.~\cite{capozzi2019dune}. 
Values of $E_i$ are obtained from PE using the inverse of Eq.~\ref{eq:CarlsModel}.  
Limits are obtained for each of the four signal models described previously: flat, power law, supernova-like, and mono-energetic.
In the mono-energetic case, \mbox{$\phi(E_i) = 1$}, i.e. for each value of $E_i$ considered, all the neutrinos are assumed to have that energy. 
For each merger, in each energy bin the method described by Feldman and Cousins~\cite{feldman} is used to construct confidence intervals on the number of signal events that could have been measured, and from these confidence intervals the values of $N_{\rm limit}(E_i)$ are determined.

\begin{figure}[!t]
     \centering
     \includegraphics[width=0.99\linewidth]{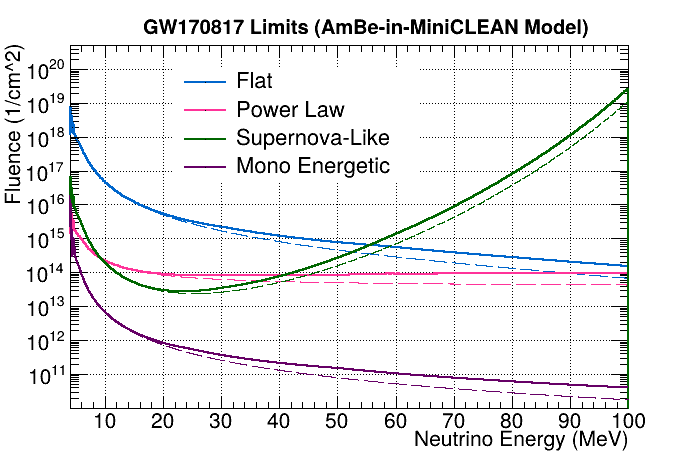}
     \includegraphics[width=0.99\linewidth]{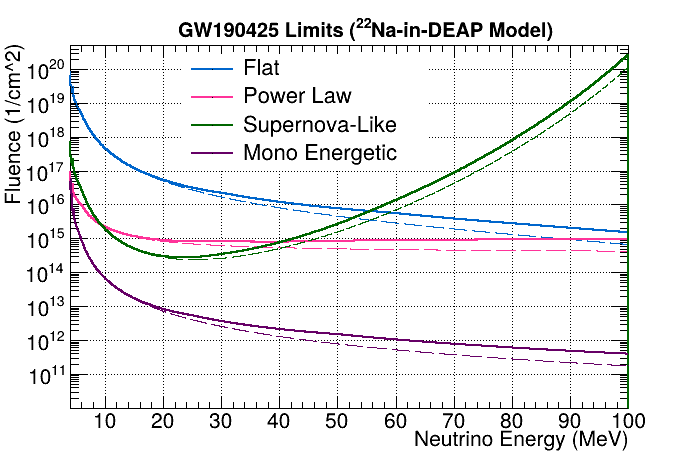}
     \includegraphics[width=0.99\linewidth]{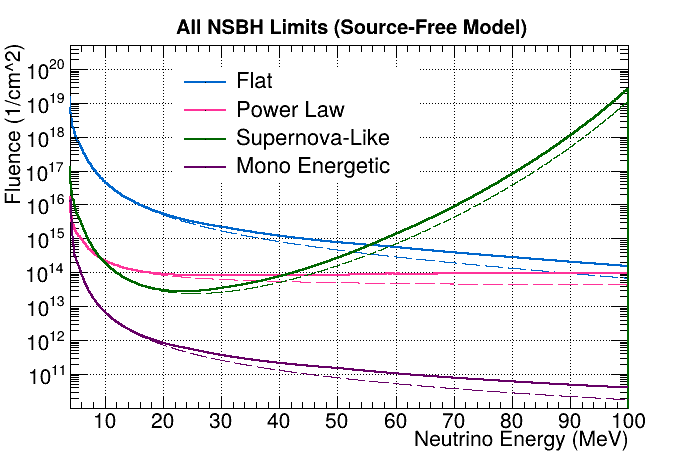}
     \caption{Exclusion limits on electron neutrino fluence from GW170817 (top), GW190425 (middle), and five NSBH mergers (bottom) for a flat, power law, supernova-like, and mono-energetic neutrino signal model. Solid and dashed lines represent the conservative and optimal signal acceptance extrapolations, respectively. }
     \label{fig:uniLimits}
\end{figure}

\begin{figure*}[!t]
\begin{minipage}[t]{0.475\linewidth}
\centering
\includegraphics[width=\textwidth]{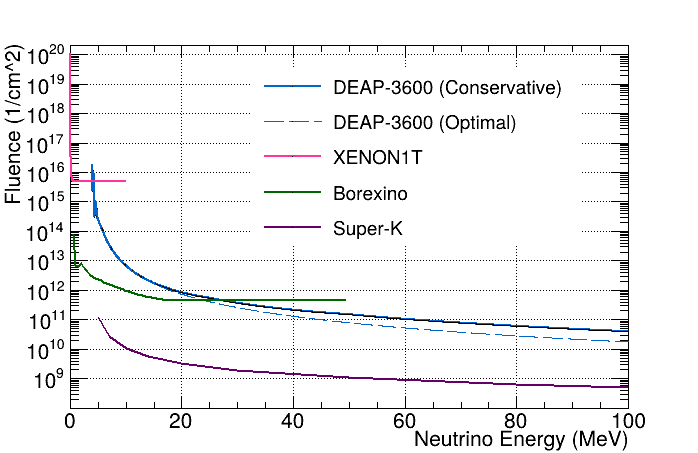}
\caption{Upper limits on mono-energetic electron neutrino fluence from NSNS merger GW170817 as set by this work, XENON1T~\cite{xenon}, Borexino~\cite{borexino}, and Super-Kamiokande~\cite{superk-170817}.}
\label{fig:result170817}
\end{minipage}
\hspace{0.5cm}
\begin{minipage}[t]{0.475\linewidth}
\centering
\includegraphics[width=\textwidth]{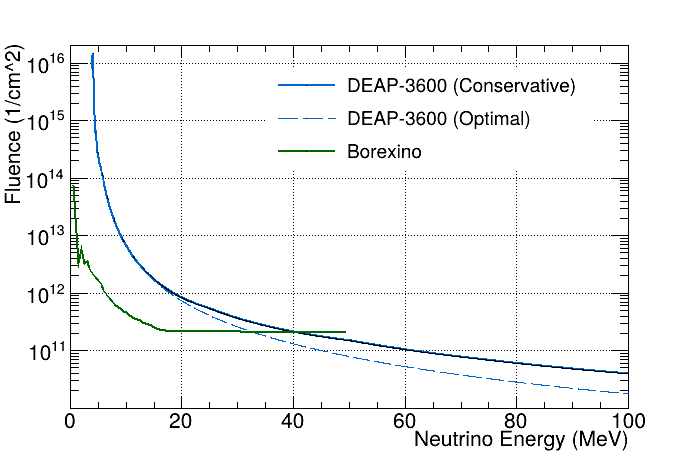}
\caption{Upper limits on mono-energetic electron neutrino fluence from NSBH mergers as set by this work and Borexino~\cite{borexino}.}
\label{fig:resultNSBH}
\end{minipage}
\end{figure*}

\begin{table}[htb]
    \centering
    \caption{The NLL values of the data for each merger against its background model, and the models' $\alpha$ values corresponding to a 90\% confidence level. To reject the background model, a merger's NLL value must be greater than $\alpha$.}
    \begin{tabular}{c c c c}
        \hline
        \textbf{Merger} & \textbf{Background Model \  } & \textbf{NLL$_{Data}$} & \textbf{\ \ $\alpha$\ \ } \\
        \hline
        \multicolumn{1}{l}{GW170817} & AmBe-in-MiniCLEAN  \  & 35.3 & 42.7 \\
        \multicolumn{1}{l}{GW190425} & $^{22}$Na-in-DEAP  \  & 17.1 & 21.3 \\
        \multicolumn{1}{l}{GW190426\_152155} & Source-Free  \  & 35.2 & 42.3 \\
        \multicolumn{1}{l}{GW190917} & Source-Free  \  & 35.4 & 42.3 \\
        \multicolumn{1}{l}{GW200105} & Source-Free  \  & 38.4 & 42.3 \\
        \multicolumn{1}{l}{GW200115} & Source-Free  \  & 33.2 & 42.3 \\
        \multicolumn{1}{l}{GW200210} & Source-Free  \  & 32.5 & 42.3 \\
        \hline
    \end{tabular}
    \label{tab:NLL}
\end{table}

Figure~\ref{fig:uniLimits} shows the exclusion limits for all four signal models for GW170817, GW190425, and all five NSBH mergers. For the NSBH mergers, limits are drawn for each signal type utilizing the average values of $N_{\rm limit}$, while limits for the two NSNS mergers with unique background models are drawn separately. The most restrictive upper limit for each merger is the mono-energetic case, since a sharp peak in energy of neutrinos is easily distinguished from backgrounds with relatively few signal events across the models: 
for example, in the case of GW170817, for 10~MeV (100~MeV) electron neutrinos these limits are set at 
6.8$\times10^{12}~\text{cm$^{-2}$}$ (1.8$\times10^{10}~\text{cm$^{-2}$}$). 
The limits for GW190425 are weaker for all signal models due to the effects of prescaling having limited the fraction of events recorded during this merger.

The upper limits on the mono-energetic electron neutrino fluence from NSNS merger GW170817 is compared to the corresponding limits from other experiments in Figure~\ref{fig:result170817}. 
Limits on the mono-energetic electron neutrino fluence from the NSBH mergers are compared to the corresponding limits from Borexino~\cite{borexino} in Figure~\ref{fig:resultNSBH}. 
Although the number of targets in DEAP is significantly smaller ($^{40}$Ar nuclei in DEAP vs. electrons in Borexino) the charged current neutrino absorption cross section on $^{40}$Ar is orders of magnitude higher than the cross section for electron neutrino elastic scattering on electrons.

\section{Conclusion}
Here we have performed the first multimessenger data analysis in an LAr target looking for neutrinos from mergers involving neutron stars. The analysis utilized empirical background models from sideband data to search for electron neutrino absorption events during a 1000-second window around the arrival of a GW for seven different mergers. No statistically significant evidence for signal events was found using either a cut-and-count approach or an NLL goodness-of-fit test. 

We set limits on the fluence of electron neutrinos above 3.83~MeV in energy reaching DEAP-3600 for four hypothetical neutrino signal models, the most stringent of which are obtained in the mono-energetic model. 
The limit corresponding to merger GW170817, is compared to previous results from a noble liquid scintillation detector, namely XENON1T~\cite{xenon}, and those set by larger scintillator experiments Borexino~\cite{borexino} and Super-Kamiokande~\cite{superk-170817}. Future low-background LAr detectors currently under construction such as DarkSide-20k~\cite{Aalsethetal2018} and DUNE~\cite{capozzi2019dune} may provide the increase in sensitivity necessary to finally capture this elusive signal should more mergers be detected in their livetimes. 
DEAP-3600 itself may be able to contribute more to this effort in the future by using data taken in its upgraded detector configuration. This overlaps with a planned LVK run, both of which could be coincident with new mergers involving neutron stars.

\begin{acknowledgments}
We thank the Natural Sciences and Engineering Research Council of Canada (NSERC),
the Canada Foundation for Innovation (CFI),
the Ontario Ministry of Research and Innovation (MRI), 
and Alberta Advanced Education and Technology (ASRIP),
the University of Alberta,
Carleton University, 
Queen's University,
the Canada First Research Excellence Fund through the Arthur B.~McDonald Canadian Astroparticle Physics Research Institute,
the U.S. National Science Foundation under Awards No. 2310049 and 2310091,
SECIHTI Project No. CBF-2025-I-1589,
DGAPA UNAM Grant No. PAPIIT IN102326,
the European Research Council Project (ERC StG 279980),
the UK Science and Technology Facilities Council (STFC) (ST/K002570/1 and ST/R002908/1),
the Leverhulme Trust (ECF-20130496),
the Russian Science Foundation (Grant No. 21-72-10065),
the Spanish Ministry of Science and Innovation (PID2022-138357NB-C22),
the National Science Centre, Poland (2022/47/B/ST2/02015),
the International Research Agenda Programmes of the Foundation for Polish Science: AstroCeNT (MAB/2018/7), funded from the European Regional Development Fund, 
and Astrocent (FENG.02.01-IP.05-A015/25), co-financed by the European Union under the European Funds for Smart Economy 2021-2027 (FENG); 
Teaming for Excellence grant Astrocent Plus (101137080) funded by the European Union
with complementary national funding from the Polish Ministry of Science and Higher Education (MNiSW/2025/DIR/811).
Studentship support from
the Rutherford Appleton Laboratory Particle Physics Division,
STFC and SEPNet PhD is acknowledged.
We thank SNOLAB and its staff for support through underground space, logistical, and technical services.
SNOLAB operations are supported by the CFI
and Province of Ontario MRI,
with underground access provided by Vale at the Creighton mine site.
We thank Vale for their continuing support, including the work of shipping the acrylic vessel underground.
We gratefully acknowledge the support of the Digital Research Alliance of Canada,
Calcul Qu\'ebec,
the Centre for Advanced Computing at Queen's University,
the Computational Centre for Particle and Astrophysics (C2PAP) at the Leibniz Supercomputer Centre (LRZ)
and SNOLAB
for providing the computing resources required to undertake this work.
\end{acknowledgments}

\bibliography{References}

@article{PhysRevD.34.2088,
  title = {{Inverse ${\ensuremath{\beta}}^{\mathrm{\ensuremath{-}}}$ decay of $^{40}\mathrm{Ar}$: A new approach for observing MeV neutrinos from laboratory and astrophysical sources}},
  author = {Raghavan, R. S.},
  journal = {Phys. Rev. D},
  volume = {34},
  issue = {7},
  pages = {2088--2091},
  numpages = {0},
  year = {1986},
  month = {Oct},
  publisher = {American Physical Society},
  doi = {10.1103/PhysRevD.34.2088},
  url = {https://link.aps.org/doi/10.1103/PhysRevD.34.2088}
}

@article{PhysRevC.80.055501,
  title = {Weak-interaction strength from charge-exchange reactions versus $\ensuremath{\beta}$ decay in the $A=40$ isoquintet},
  author = {Bhattacharya, M. and Goodman, C. D. and Garc\'ia, A.},
  journal = {Phys. Rev. C},
  volume = {80},
  issue = {5},
  pages = {055501},
  numpages = {6},
  year = {2009},
  month = {Nov},
  publisher = {American Physical Society},
  doi = {10.1103/PhysRevC.80.055501},
  url = {https://link.aps.org/doi/10.1103/PhysRevC.80.055501}
}

@article{GARDINER2021108123,
title = {{Simulating low-energy neutrino interactions with MARLEY}},
journal = {Comput. Phys. Commun.},
volume = {269},
pages = {108123},
year = {2021},
issn = {0010-4655},
doi = {https://doi.org/10.1016/j.cpc.2021.108123},
url = {https://www.sciencedirect.com/science/article/pii/S0010465521002356},
author = {Steven Gardiner}
}

@article{Gardiner_PhysRevC.103.044604,
  title = {Nuclear de-excitations in low-energy charged-current ${\ensuremath{\nu}}_{e}$ scattering on $^{40}\mathrm{Ar}$},
  author = {Gardiner, Steven},
  journal = {Phys. Rev. C},
  volume = {103},
  issue = {4},
  pages = {044604},
  numpages = {22},
  year = {2021},
  month = {Apr},
  publisher = {American Physical Society},
  doi = {10.1103/PhysRevC.103.044604},
  url = {https://link.aps.org/doi/10.1103/PhysRevC.103.044604}
}

@article{deap3600_design,
    author = {P.-A. Amaudruz and others},
    collaboration = {DEAP Collaboration},
    title = {{Design and construction of the DEAP-3600 dark matter detector}},
      journal = {Astropart. Phys.},
         year = 2019,
        month = mar,
       volume = {108},
        pages = {1-23},
          doi = {10.1016/j.astropartphys.2018.09.006},
}

@article{Adhikari:2023ces,
    author = "Adhikari, P. and others",
    collaboration = {DEAP Collaboration},
    title = {Precision Measurement of the Specific Activity of 39{Ar} in Atmospheric Argon with the {DEAP-3600} Detector},
	volume = {83},
	url = {https://doi.org/10.1140/epjc/s10052-023-11678-6},
	doi = {10.1140/epjc/s10052-023-11678-6},
	pages = {642},
	journal = {Eur. Phys. J. C},
    year = "2023"
}

@article{adhikari_pulse-shape_2021,
	title = {Pulse-shape discrimination against low-energy {Ar}-39 beta decays in liquid argon with 4.5 tonne-years of {DEAP}-3600 data},
	volume = {81},
	url = {https://doi.org/10.1140/epjc/s10052-021-09514-w},
	doi = {10.1140/epjc/s10052-021-09514-w},
	number = {9},
	journal = {Eur. Phys. J. C},
	author = {Adhikari, P. and others},
    collaboration = {DEAP Collaboration},
	month = sep,
	year = {2021},
	pages = {823}
}

@phdthesis{Rethmeier:2021oez,
    author = "Rethmeier, Carl",
    title = "{Characterization of Alpha Decays and Detector Response and Search for 5.5 MeV Solar Axions in DEAP-3600}",
    doi = "10.22215/etd/2021-14674",
    school = "Carleton U.",
    year = "2021"
}

@misc{RAT,
	author = {Bolton, T. and others},
	title = {{RAT (is an Analysis Tool) User's Guide}},
	year = {2018},
	url = {https://rat.readthedocs.io}
}

@article{Aalsethetal2018,
title={{DarkSide-20k}: A 20 tonne two-phase {LAr TPC} for direct dark matter detection at {LNGS}},
volume={133},
number={3},
journal={Eur. Phys. J. Plus},
author={Aalseth, C.~E. and others},
collaboration = {DarkSide-20k Collaboration},
year={2018},
month={Mar},
url={https://link.springer.com/article/10.1140/epjp/i2018-11973-4}
}

@article{capozzi2019dune,
  title={{DUNE} as the next-generation solar neutrino experiment},
  author={Capozzi, Francesco and Li, Shirley Weishi and Zhu, Guanying and Beacom, John F},
  DOI={https://doi.org/10.1103/PhysRevLett.123.131803},
  journal={Phys. Rev. Lett.},
  volume={123},
  number={13},
  pages={131803},
  year={2019},
  publisher={APS}
}

@article{PhysRevC.58.3677,
  title = {Neutrino absorption efficiency of an ${}^{40}\mathrm{Ar}$ detector from the \ensuremath{\beta} decay of ${}^{40}\mathrm{Ti}$},
  author = {Bhattacharya, M. and others},
  journal = {Phys. Rev. C},
  volume = {58},
  issue = {6},
  pages = {3677--3687},
  numpages = {0},
  year = {1998},
  month = {Dec},
  publisher = {American Physical Society},
  doi = {10.1103/PhysRevC.58.3677},
  url = {https://link.aps.org/doi/10.1103/PhysRevC.58.3677}
}

@phdthesis{AndrewThesis,
    author = "Erlandson, Andrew",
    title = "{First observation of solar neutrino absorption on \textsuperscript{40}Ar using the DEAP-3600 detector}",
    doi = "10.22215/etd/2024-16484",
    school = "Carleton U.",
    year = "2024"
}

@article{PhysRevD.100.072009,
  title = {{Electromagnetic backgrounds and potassium-42 activity in the DEAP-3600 dark matter detector}},
  author = {Ajaj, R. and others},
  collaboration = {DEAP Collaboration},
  journal = {Phys. Rev. D},
  volume = {100},
  issue = {7},
  pages = {072009},
  numpages = {17},
  year = {2019},
  month = {Oct},
  publisher = {American Physical Society},
  doi = {10.1103/PhysRevD.100.072009},
  url = {https://link.aps.org/doi/10.1103/PhysRevD.100.072009}
}

@article{antares,
    author = {A. Albert and others},
    collaboration = {ANTARES Collaboration},
    title = {{Search for neutrino counterparts to the gravitational wave sources from LIGO-Virgo O3 run with the ANTARES detector}},
    journal = {JCAP},
    volume = {4},
    year = 2023,
    url = {https://doi.org/10.1088/1475-7516/2023/04/004},
    pages = {13}
}

@article{borexino,
    author = {D. Basilico and others},
    collaboration = {Borexino Collaboration},
    title = {{Borexino's search for low-energy neutrinos associated with gravitational wave events from GWTC-3 database}},
    journal = {Eur. Phys. J. C},
    volume = {83},
    year = 2023,
    url = {https://doi.org/10.1140/epjc/s10052-023-11688-4},
    pages = {13}
}

@conference{icecube,
    title = {{First results of low-energy neutrino follow-ups of Run O4 compact binary mergers with the IceCube Neutrino Observatory}},
    author = {R. Abbasi and others},
    collaboration = {IceCube Collaboration},
    booktitle = {The 38th International Cosmic Ray Conference},
    organization = {Nagoya, Japan},
    year = 2023
}

@article{km3net,
    author = {S. Aiello and others},
    collaboration = {KM3NeT Collaboration},
    title = {{Searches for neutrino counterparts of gravitational waves from the LIGO/Virgo third observing run with KM3NeT}},
    journal = {JCAP},
    volume = {4},
    year = 2024,
    url = {https://doi.org/10.1088/1475-7516/2024/04/026},
    pages = {19}
}

@article{xenon,
    author = {E. Aprile and others},
    collaboration = {XENON Collaboration},
    title = {{Search for events in XENON1T associated with gravitational waves}},
    journal = {Phys. Rev. D},
    volume = {108},
    year = 2023,
    url = {https://doi.org/10.1103/PhysRevD.108.072015},
    pages = {9}
}

@article{superk,
    author = {K. Abe and others},
    collaboration = {Super-Kamiokande Collaboration},
    title = {{Search for neutrinos in coincidence with gravitational wave events from the LIGO-Virgo O3a observing run with the Super-Kamiokande detector}},
    journal = {ApJL},
    volume = {918},
    year = 2021,
    url = {https://doi.org/10.3847/1538-4357/ac0d5a},
    pages = {16}
}

@article{superk-170817,
    author = {K. Abe and others},
    collaboration = {Super-Kamiokande Collaboration},
    title = {{Search for neutrinos in Super-Kamiokande associated with the GW170817 neutron-star merger}},
    journal = {ApJL},
    volume = {857},
    year = {2018},
    url = {https://doi.org/10.3847/2041-8213/aabaca},
    pages = {8}
}

@article{eichler,
    author = {Eichler, David and Livio, Mario and Piran, Tsvi},
    title = {{Nucleosynthesis, neutrino bursts, and gamma-rays from coalescing neutron stars}},
    journal = {Nature},
    volume = {340},
    year = 1989,
    url = {https://doi.org/10.1038/340126a0},
    pages = {7}
}

@article{ardevol,
    author = {Ardevol-Pulpillo, R. and Janka, H.T. and Just, O. and Bauswein, A.},
    title = {{Improved leakage-equilibration-absorption scheme (ILEAS) for neutrino physics in compact object mergers}},
    journal = {Mon. Not. R. Astron. Soc.},
    volume = {485},
    year = 2019,
    url = {https://doi.org/10.1093/mnras/stz613},
    pages = {39}
}

@article{burns,
    author = {Burns, Eric},
    title = {Neutron star mergers and how to study them},
    journal = {Living Rev. Relativ.},
    volume = {23},
    year = 2020,
    url = {https://doi.org/10.1007/s41114-020-00028-7},
    pages = {177}
}

@article{foucart,
    author = {Foucart, Francois},
    title = {Neutrino transport in general relativistic neutron star merger simulations},
    journal = {Living Rev. Comput. Astrophys.},
    volume = {9},
    year = 2023,
    url = {https://doi.org/10.1007/s41115-023-00016-y},
    pages = {97}
}

@article{kimuraNSNS,
    author = {Kimura, S. and others},
    title = {Trans-ejecta high-energy neutrino emission from binary neutron star mergers},
    journal = {Phys. Rev. D},
    volume = {98},
    year = 2018,
    url = {https://doi.org/10.1103/PhysRevD.98.043020},
    pages = {12}
}

@article{tamborra,
    author = {Irene Tamborra},
    title = {Neutrinos from explosive transients at the dawn of multi-messenger astronomy},
    journal = {Nat. Rev. Phys.},
    volume = {7},
    year = 2025,
    url = {https://doi.org/10.1038/s42254-025-00828-2},
    pages = {23}
}

@article{sn1987nu,
    author = {K. Hirata and others},
    title = {{Observation of a neutrino burst from the supernova SN1987A}},
    journal = {Phys. Rev. Lett.},
    volume = {58},
    year = 1987,
    url = {https://doi.org/10.1103/PhysRevLett.58.1490},
    pages = {4}
}

@article{firstGW,
    author = {B. P. Abbott and others},
    collaboration = {LIGO Scientific Collaboration and VIRGO Collaboration},
    title = {Observation of Gravitational Waves from a Binary Black Hole Merger},
    journal = {Phy. Rev. Lett.},
    volume = {116},
    year = 2016,
    url = {https://doi.org/10.1103/PhysRevLett.116.061102},
    pages = {16}
}

@article{LIGO-1,
    author = {B. P. Abbott and others},
    collaboration = {LIGO Scientific Collaboration and VIRGO Collaboration},
    title = {{GWTC-1: A gravitational-wave transient catalog of compact binary mergers observed by LIGO and Virgo during the first and second observing runs}},
    journal = {Phys. Rev. X},
    volume = {9},
    year = 2019,
    url = {https://doi.org/10.1103/PhysRevX.9.031040},
    pages = {49}
}

@article{LIGO-2,
    author = {R. Abbott and others},
    collaboration = {LIGO Scientific Collaboration and VIRGO Collaboration},
    title = {{GWTC-2: Compact binary coalescences observed by LIGO and Virgo during the first half of the third observing run}},
    journal = {Phys. Rev. X},
    volume = {11},
    year = 2021,
    url = {https://doi.org/10.1103/PhysRevX.11.021053},
    pages = {54}
}

@article{LIGO-2.1,
    author = {R. Abbott and others},
    collaboration = {LIGO Scientific Collaboration and VIRGO Collaboration},
    title = {{GWTC-2.1: Deep extended catalog of compact binary coalescences observed by LIGO and Virgo during the first half of the third observing run}},
    journal = {Phys. Rev. D},
    volume = {109},
    year = 2024,
    url = {https://doi.org/10.1103/PhysRevD.109.022001},
    pages = {46}
}

@article{LIGO-3,
    author = {R. Abbott and others},
    collaboration = {LIGO Scientific Collaboration, VIRGO Collaboration, and KAGRA Collaboration},
    title = {{GWTC-3: Compact binary coalescences observed by LIGO and Virgo during the second part of the third observing run}},
    journal = {Phys. Rev. X},
    volume = {13},
    year = 2023,
    url = {https://doi.org/10.1103/PhysRevX.13.041039},
    pages = {88}
}

@article{nuAbsPaper,
    author = {P. Adhikari and others},
    collaboration = {DEAP Collaboration},
    title = {{First evidence of neutrino absorption on argon using $^8$B solar neutrinos in DEAP-3600}},
    journal = {\textit{Preprint} arXiv},
    volume = {2605.12769},
    year = {2026},
    url = {https://doi.org/10.48550/arXiv.2605.12769},
    pages = {6}
}

@article{deap_39Ar_halflife,
	title = {{Direct measurement of the 39Ar half-life from 3.4 years of data with the DEAP-3600 detector}},
	author = {P. Adhikari and others},
	collaboration = {{DEAP-3600 Collaboration}},
	year = {2025},
	journal = {Eur. Phys. J. C},
	volume = {85},
	pages = {728},
	publisher = {{Springer Berlin Heidelberg}},
	doi = {10.1140/epjc/s10052-025-14289-5},
	url = {https://doi.org/10.1140/epjc/s10052-025-14289-5}
}

@article{baret,
    author = {B. Baret and others},
    title = {Bounding the Time Delay between High-energy Neutrinos and Gravitational-wave Transients from Gamma-ray Bursts},
    journal = {Astropart. Phys.},
    volume = {35},
    year = 2011,
    url = {https://doi.org/10.1016/j.astropartphys.2011.04.001},
    pages = {23}
}

@article{berezinsky,
    author = {Berezinsky, V.S. and Prilutsky, O. F.},
    title = {Pulsars and Cosmic Rays in the Dense Supernova Shells},
    journal = {Astronomy and Astrophysics},
    volume = {66},
    year = 1978,
    pages = {10}
}

@article{pwrLaws,
    author = {M. E. J. Newman},
    title = {Power laws, {Pareto} distributions and {Zipf}'s law},
    journal = {Contemporary Physics},
    volume = {46},
    year = 2005,
    url = {https://doi.org/10.1080/00107510500052444},
    pages = {29}
}

@article{superSpectrum,
    author = {Huang, Xu-Run and Sun, Chuan-Le and Chen, Lie-Wen and Gao, Jun},
    title = {Bayesian Inference of Supernova Neutrino Spectra with Multiple Detectors},
    journal = {JCAP},
    volume = {9},
    year = 2023,
    url = {https://doi.org/10.48550/arXiv.2305.00392},
    pages = {25} 
}

@article{swope,
    author = {Coulter, D. A. and Foley, R.J and Kilpatrick, C. D. and Drout, M. R and Piro, A. L.},
    title = {{Swope Supernova Survey 2017a (SSS17a), the optical counterpart to a gravitational wave source}},
    journal = {Science},
    volume = {358},
    year = 2017,
    url = {https://doi.org/10.1126/science.aap9811},
    pages = {25}
}

@article{gw190425,
    author = {Romero-Shaw, I. M. and Farrow, N. and Stevenson, S. and Thrane, E. and Zhu, X.},
    title = {{On the origin of GW190425}},
    journal = {Mon. Not. Roy. Astron. Soc.},
    year = {2020},
    volume = {496},
    url = {https://doi.org/10.1093/mnrasl/slaa084},
    pages = {5}
}

@article{gw200210,
    author = {Du, Z. and L{\"u}, H. and Liu, X. and Fan, X. and Liang, E.},
    title = {{How to identify the object with mass range of (2.2−3) M$_{\odot}$ in the merger of compact star systems}},
    journal = {Astrophys.J.},
    year = 2025,
    volume = 985,
    url = {https://doi.org/10.3847/1538-4357/adcb49},
    pages = {16}
}

@article{miniclean,
    author = {Rielage, K. and others},
    title = {{Update on the MiniCLEAN dark matter experiment}},
    journal = {Physics Procedia},
    year = {2015},
    volume = {61},
    url = {https://doi.org/10.1016/j.phpro.2014.12.024},
    pages = {8}
}

@article{feldman,
    author = {Feldman, Gary J. and Cousins, Robert D.},
    title = {Unified approach to the classical statistical analysis of small signals},
    journal = {Phys. Rev. D},
    volume = {57},
    year = 1998,
    url = {https://doi.org/10.1103/PhysRevD.57.3873},
    pages = {40}
}

\end{document}